\documentclass{article}
\usepackage{spconf,amsmath,graphicx}
\usepackage{color}
\usepackage{url}
\usepackage{graphicx}
\title{Residual-guided Personalized Speech Synthesis Based on Face Image}
\name{Jianrong Wang$^1$, Zixuan Wang$^1$, Xiaosheng Hu$^1$, Xuewei Li$^1$, Qiang Fang$^2$, Li Liu$^{3,*}$\thanks{* Corresponding author}}
\address{$^1$College of Intelligence and Computing, Tianjin University, Tianjin, China\\
$^2$Institute of Linguistics, Chinese Academy of Social Sciences, Beijing, China\\
$^3$Shenzhen Research Institute of Big Data, the Chinese University of Hong Kong, Shenzhen, China}
\begin{document}
%\ninept
%
\maketitle
\vspace{-0.1cm}
\begin{abstract}
Previous works derive personalized speech features by training the model on a large dataset composed of his/her audio sounds. It was reported that face information has a strong link with the speech sound. Thus in this work, we innovatively extract personalized speech features from human faces to synthesize personalized speech using neural vocoder. A \textbf{F}ace-based \textbf{R}esidual \textbf{P}ersonalized \textbf{S}peech \textbf{S}ynthesis Model (FR-PSS) containing a speech encoder, a speech synthesizer and a face encoder is designed for PSS. In this model, by designing two speech priors, a residual-guided strategy is introduced to guide the face feature to approach the true speech feature in the training. Moreover, considering the error of feature's absolute values and their directional bias, we formulate a novel tri-item loss function for face encoder. Experimental results show that the speech synthesized by our model is comparable to the personalized speech synthesized by training a large amount of audio data in previous works.
\end{abstract}
\vspace{-0.1cm}
\section{Introduction}
\vspace{-0.1cm}
%Concerning the feasibility of generating speech from face, some previous studies have shown that the voice of a person is strongly related to his/her facial structures \cite{KAMACHI20031709}. For example, it was shown in \cite{teager1990evidence} that facial bone, joint structures and the tissues covering them are closely related to the shape and size of the organs that produce sound. Meanwhile, genetic factors, biological factors and environmental factors, especially gender \cite{inproceedings}, age \cite{Zazo2018Age} and ethnicity can largely influence one's voice and face.

%Recently, some research works on speech to face generation using deep learning methods have emerged \cite{Oh_2019_CVPR,wang2020attentionbased}, they fed speech to a Convolutional Neural Network (CNN) \cite{NIPS2012_c399862d} to learn facial features \cite{merler2019diversity}, and then used these features to generate face image \cite{Cole_2017_CVPR}. Another method was based on the encoder-decoder structure \cite{Oh_2019_CVPR,wang2020attentionbased} to generate face images from speech. Motivated by these promising results, we believe that the inverse direction, \textit{i.e.}, the personalized speech synthesis from one's face image is feasible.

Generating natural speech from text (text-to-speech synthesis, TTS) has been studied for decades \cite{wang2017tacotron,8461368,liu2018visual}. Tacotron \cite{wang2017tacotron} and Tacotron 2 \cite{8461368} are two efficient end-to-end speech synthesis models for TTS based on deep learning. As for the personalized speech synthesis of multiple speakers, Google proposed a system \cite{jia2018transfer} that used several audio datasets to train a speaker encoder with a sequence-to-sequence TTS network and the vocoder based on Tacotron2. Besides, Baidu’s deepvoice3 \cite{DBLP:journals/corr/abs-1710-07654} added a speaker encoder on the basis of Tacotron to characterize the timbre characteristics of speakers.

%As for the personalized speech synthesis of multiple speakers, a neural network-based system for multi-speaker TTS synthesis was proposed in \cite{jia2019transfer} by Google. The system used several audio datasets to train a speaker encoder network with a sequence-to-sequence TTS synthesis network and the vocoder based on Tacotron 2. Besides, Baidu's deepvoice3 \cite{DBLP:journals/corr/abs-1710-07654} learned 2500 voices in half an hour, and a speaker encoder was added on the basis of Tacotron \cite{wang2017tacotron} to characterize the timbre characteristics of speakers.

Most of current works extracted personalized speech features based on a large volume of audio signals. Recently, some research works on speech to face generation using deep learning methods have emerged \cite{Oh_2019_CVPR,wang2020attentionbased,yi2020audio}. Motivated by these promising results, we believe that the inverse direction, \textit{i.e.}, the personalized speech synthesis from one's face image is feasible. Face2Speech \cite{goto2020face2speech} is the method to predict speech features from face images and then synthesize speech using the WORLD vocoder \cite{morise2016world}. However, it was reported in \cite{DBLP:journals/corr/abs-1710-07654} that WORLD vocoder introduces various noticeable artifacts, and the WaveNet vocoder sounds more natural. Besides, \cite{goto2020face2speech} only performed qualitative experiments without quantitative experiments, thus lacking objective evaluation criteria.

In this work, we propose a novel speaker-independent \textit{Face-based Residual Personalized Speech Synthesis Model (FR-PSS)} within an encoder-decoder architecture, which extracts personalized information from the face image, and synthesizes audio speech using natural vocoder. An overview of our proposed FR-PSS is shown in Fig. \ref{fig:pipeline}. Firstly, the speech encoder is trained to extract features from speech. Secondly, the speech synthesizer is trained to synthesize speech from a given text and features generated from the pre-trained speech encoder. Thirdly, the face encoder is trained with the pairs of  face image and speech for making features extracted from a face image of a speaker closer to one derived from his/her speech. The tri-item loss function and a residual-guided strategy is introduced in this step. Finally, the face encoder and speech synthesizer are concatenated for building our FR-PSS model. Compared with Face2Speech, we conduct a series of quantitative experiments. Experimental results show that our FR-PSS achieves a comparable performance compared with method that synthesizes personalized speech using a speech-derived embedding vector. 

Overall, our contributions can be summarized as follows. i) A new speech synthesis model FR-PSS is proposed. We extract personalized speech features from human faces to synthesize personalized speech using neural vocoder. ii) A residual-guided strategy is designed by incorporating a prior speech feature to make the network capture more representative face features and improve model learning efficiency. iii) We innovatively establish a tri-item loss function to accelerate training convergence.
\vspace{-0.1cm}
\section{Face-based Residual Personalized Speech Synthesis Model}
\vspace{-0.1cm}
\begin{figure*}[htbp]
	\centering
	\includegraphics[width=0.83\textwidth,height=2.5in]{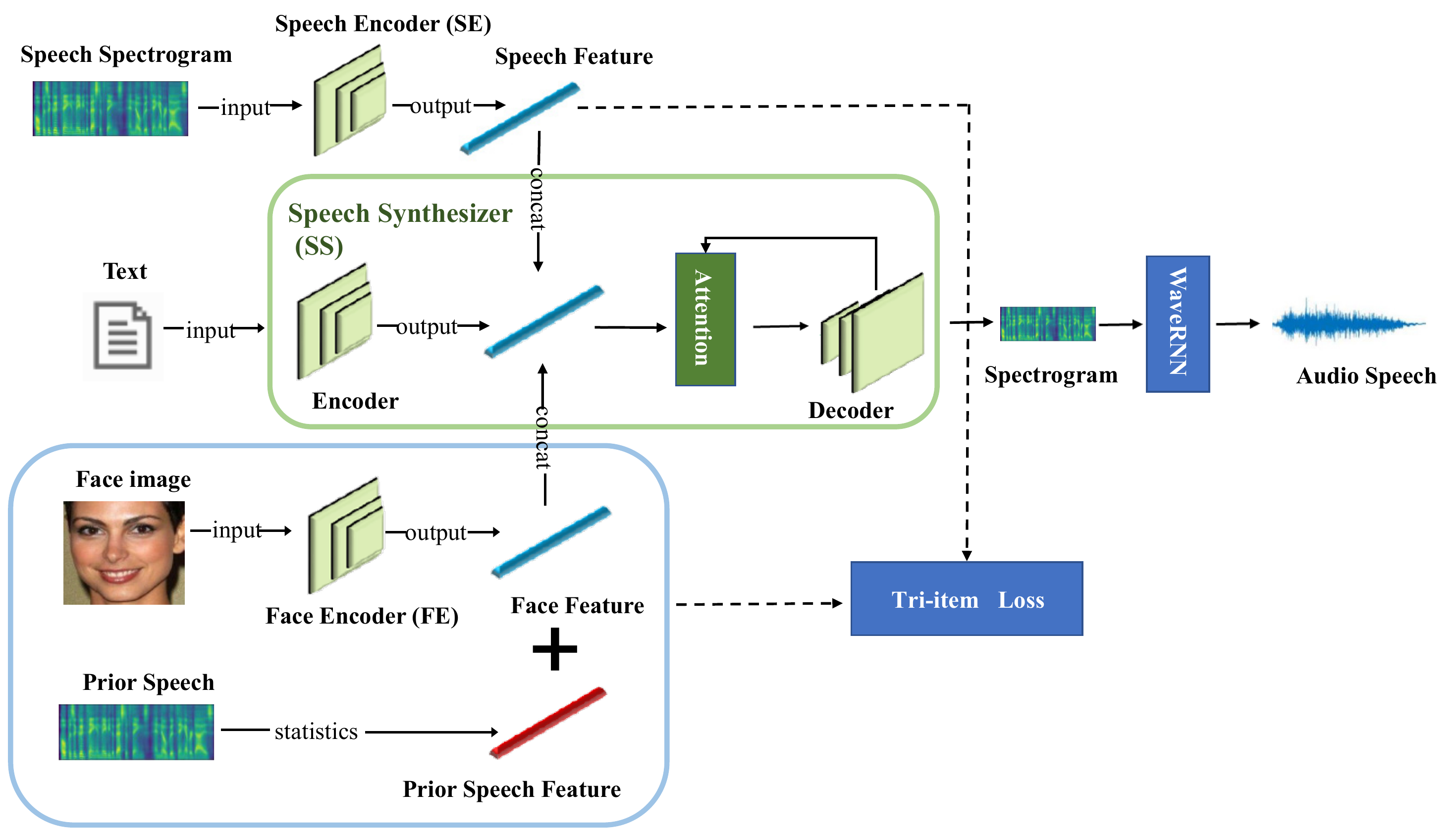}
	\caption{Overview of our FR-PSS with the prior speech.}
	\label{fig:pipeline}
\vspace{-0.3cm}
\end{figure*}

In this section, we will introduce the proposed FR-PSS (shown in Fig. \ref{fig:pipeline}), which includes Speech Encoder (\textit{SE}), Speech Synthesizer (\textit{SS}), Face Encoder (\textit{FE}), Residual-guided strategy and tri-item loss function, respectively.
\vspace{-0.2cm}
\subsection{Speech Encoder and Speech Synthesizer}

\textit{SE} is to extract speech features. Similar to \cite{2018Generalized}, we apply a network that maps a sequence of Mel spectrogram frames calculated from speech to a fixed-dimensional embedding vector. The 40-channel Mel spectrogram is fed to the network, which is composed of three LSTM layers with a total of 768 units, and after each layer, there is a 256-dimensional projection. In the inference process, each utterance of speakers is divided into 800ms windows with 50\% overlapping. The network runs independently on each window, and generates the final speech embedding by the average and normalization. We use generalized end-to-end speaker verification losses to train the network.

As for the \textit{SS}, we use Tacotron 2 model, which includes an encoder and a decoder that introduces an location sensetive attention. The details can be referred to \cite{8461368}. The pretrained WaveRNN in \cite{pmlr-v80-kalchbrenner18a} is exploited as the vocoder in this work.

\vspace{-0.3cm}
\subsection{Face Encoder}

We use \textit{FE} to extract face features which is used to synthesize speech. It is a four convolution blocks structure, which stacks the convolution kernel of size 1$\times$1, 3$\times$3 and pooling operations. The dimension of the output feature maps after these three operations are the same, so that we can add them in the channel dimension. This can increase the network's width and improve its adaptability to scale. ReLU operation is performed after each convolution layer to increase the non-linear character of the network. Finally, a 256-dimensional embedding is achieved by using a 1$\times$1 convolution.

To make \textit{FE} owes an ability to focus on face features and ignore the background noise, we utilize a lightweight general purpose module CBAM \cite{Woo_2018_ECCV}, which can be easily integrated into an end-to-end CNN architecture. CBAM contains a channel and a spatial attention, which are embedded into decoder. Besides, we also tried channel-focused attention mechanism \cite{2019Squeeze}, and it turns out that CBAM performs better.
\vspace{-0.4cm}
\subsection{Residual-guided Strategy}
\vspace{-0.1cm}
Due to natural mismatch between face image and speech, it is hard to extract complete speech features from face image. Therefore, we add a prior information to help \textit{FE} extract speech features. By introducing the prior speech feature, we exploit the idea of the residual to remove the main similar part of the speech (\textit{i.e.}, prior speech feature),
thereby highlighting subtle changes depicted by speech feature. It is to reduce the training difficulties and learn more representative speech features. Our FR-PSS converts the face image to speech by a network $\phi$: $\phi(f,t)=SS\left(FE(f)+t+s_{prior}\right)$, where $SS$ is the speech synthesizer, $t$ is the text, $f$ means the image of input face, and $s_{prior}$ is the prior speech feature calculated before the training stage.

Two speech priors are investigated in this work. The first one is neutral speech prior, which is the arithmetic mean of a large gender-balance speech dataset\footnote[1]{The dataset contains the same number of male and female speakers.}: $s_{prior}=\frac{1}{n}\sum_{i=1}^{n} {{SE}}(s)$,
 where $s$ denotes the audio clips and $n$ is the number of speaker. \textit{SE} is a CNN structure to extract speech feature by taking $n$ = 10, 50, 100, 500, and 1000. In this work, we finally take $n$ equals 500.

The second prior speech feature is gender-dependent prior by assigning two prior speech features to males and females, respectively. To achieve this, a robust classifier network is first needed to predict the gender based on the face image. We use a gender prediction network \cite{2015Age} to predict the gender of speakers. It is trained on the Adience dataset \cite{2014Age} and tested on VGGFace2 dataset \cite{cao2018vggface2}.
\vspace{-0.2cm}
\subsection{Tri-item Loss Function}

In order to make \textit{FE} extract the personalized facial features better, we formulate a tri-item loss function, which is composed of $L_{2}$ loss, the negative cosine similarity loss and the triplet loss \cite{Schroff_2015_CVPR}.

We consider that $L_{2}$ loss can be used to measure the error between facial features and speech features and improve the perceptual similarity at the abstract level. The negative cosine similarity is introduced to make speech features and face features vectors similar in direction. The triplet loss is to make the distance between speech feature and the corresponding face feature vectors closer, and make the distance between the speech feature vector and the irrelevant face feature vectors farther. The loss function is given as:
     \begin{small}
	\begin{equation}\label{eq:cos}
     \setlength{\abovedisplayskip}{3pt}
     \setlength{\belowdisplayskip}{3pt}
	\begin{split}
	L_{Total} &=  1 - \frac{\sum_{i=1}^{n} A_{i} \times B_{i}}{\sqrt{\sum_{i=1}^{n}A_{i}^{2}} \times \sqrt{\sum_{i=1}^{n} B_{i}^{2}}}
	+ \frac{\sum_{i=1}^{n} (A_{i} - B_{i})^2}{n}\\
	&+ max(d(A,B)-d(A,C),0),
	\end{split}
	\end{equation}
     \end{small}where $d$ denotes distance of embedding vectors, $A$ and $B$ are the speech feature and the corresponding face feature embedding vectors. $C$ is the face feature vector of the irrelevant speaker, and $n$ represents the dimension of the feature vector.
\vspace{-0.5cm}
\section{Experiment}
%\label{sec:pagestyle}
\vspace{-0.1cm}
\subsection{Datasets}

Three public datasets are used to train the proposed FR-PSS: 
%\begin{itemize}

For the \textit{SE}, we use the Voxceleb2 \cite{chung2018voxceleb2}, which contains 6000 celebrities that contribute more than one million utterances. We reduce the audio sampling rate to 16 kHz and divide it into three subsets (\textit{i.e.}, training, validation and test). Note that there is no speaker overlapped among these three subsets for all experimental setup in this work.

For the \textit{SS}, we use LibriTTS \cite{zen2019libritts} consisting of two training sets, which contains 436 hours of audio speech uttered by 1172 English speakers, with a sampling frequency of 16 kHz.

As for the \textit{FE}, we use the voice recording from the Voxceleb2 dataset and the face images from the manually filtered version of VGGFace2 \cite{cao2018vggface2} dataset. We select an intersection of the two datasets with the common identities and eliminate non-English speakers, deriving 149,354 voice recordings and 139,572 face images of 1089 subjects. The ratio of our training set to our test set is 8:2.
%\end{itemize}
\vspace{-0.3cm}
\subsection{Implementation Details}
We introduce the implementation details for \textit{SE}, \textit{SS} and \textit{FE}, respectively.
%\begin{itemize}

First, we use the Voxceleb2 dataset to train the \textit{SE}. Our model is implemented by PyTorch 1.1.0 with GPU Tesla K80, and is optimized by Adam with the learning rate of 0.001 and the exponentially decay rate of 0.9 at every 5 epochs. We train our model with 1.56M steps and batch size 64. The number of utterance is 10 per speaker.

Secondly, based on the LibriTTS \cite{zen2019libritts}, we first train the speech synthesizer network, and then train the WaveRNN \cite{pmlr-v80-kalchbrenner18a} based on the output of the speech synthesizer network. At the speech synthesizer decoder side, the ground truth is passed in rather than the predicted results. The batch size is 64, and we use the Adam optimizer with $\beta_{1} = 0.9$, $\beta_{2} = 0.999$ and $\epsilon = 10^{-6}$. The initial value of the learning rate is $10^{-3}$, and the exponential drops to $10^{-5}$ after 50,000 iterations. The pretrained WaveRNN in \cite{pmlr-v80-kalchbrenner18a} is exploited as the vocoder in this work.

Lastly, we use the spectrogram and face-image pairs to train the \textit{FE}. The \textit{FE} is optimized by Adam and the learning rate of 0.01 with the exponentially decay rate of 0.9 at every 5 epochs. We finally train our model with 50 epochs, and the batch size is 8.
%\end{itemize}

\vspace{-0.3cm}
\subsection{Evaluation Metrics}

To evaluate the quality of speech synthesized by the FR-PSS quantitatively, we compare speech features extracted from the generated audio and the original audio, using the $L_{1}$, $L_{2}$ and Cosine Similarity loss ($Cos$). When $L_{1}$ and $L_{2}$ tends to 0 or $Cos$ tends to 1, the difference between two features are similar in the direction. Besides, we rely on Mean Opinion Score (MOS) evaluations based on subjective listening tests. It is to evaluate synthesized speech in two aspects: its naturalness and similarity to real speech from the target speaker.

\vspace{-0.1cm}

\section{Result and Analysis}
\vspace{-0.1cm}
%\label{sec:typestyle}
\subsection{Ablation Experiment on Loss Function}
\begin{figure}[htbp]
	\centering
     \setlength{\abovecaptionskip}{0.5cm}
	\includegraphics[width=0.35\textwidth]{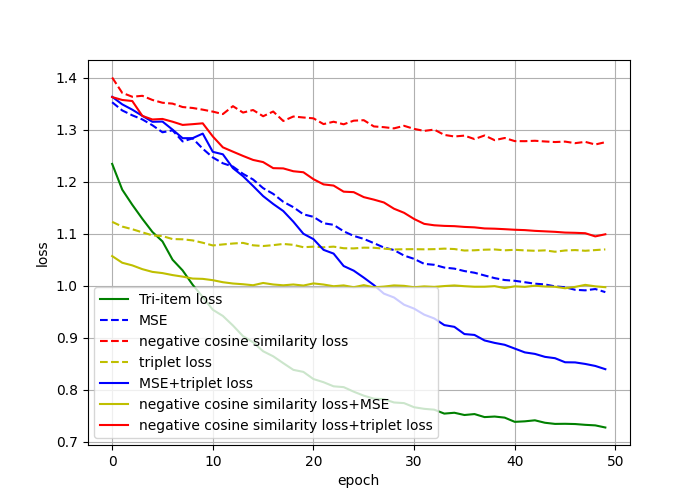}
	\caption{Convergence rate of loss functions.}
	\label{fig:loss}
\vspace{-0.3cm}
\end{figure}

To verify the effectiveness of the tri-item loss function we proposed, we compare it with different loss functions. The decline of loss functions are shown in Fig. \ref{fig:loss}. We can see that the convergence speed is obviously accelerated when the proposed tri-item loss function is used to train the model. It shows that the proposed tri-item loss function is effective.
\vspace{-0.3cm}
\vspace{-0.4cm}
\subsection{Quantitative Result}
\vspace{-0.1cm}
%Recall that in the previous work \cite{wang2017tacotron,8461368,jia2019transfer,DBLP:journals/corr/abs-1710-07654}, a set of audio data are used to train a model to obtain the personalized speech feature. The personalized speech synthesized by this approach is called $\rm Audio_{s}$ here. In Table \ref{tab:feature_similarity}, we calculate the difference between $\rm Audio_{s}$ and the true speech feature ($\rm T_{speech}$), and the difference between speech features synthesized by the FR-PSS from face image with $\rm T_{speech}$ using different types of priors. We compare their distance by the $L_{1}$, $L_{2}$ and $Cos$. It can be seen that the speech obtained by FR-PSS from face images is close to the $\rm Audio_{s}$. In particular, adding gender prior knowledge can improve performance significantly.
Recall that in the previous work \cite{wang2017tacotron,8461368,jia2018transfer,DBLP:journals/corr/abs-1710-07654}, a set of audio data are used to train a model to obtain the personalized speech feature. We call this method that synthesize personalized speech using speaker audio embeddings SYNTH-AUDIO in this work. While the embedding vector of SYNTH-AUDIO is made by applying the \textit{SE} to the audio of the speaker, that of FR-PSS is generated by applying the \textit{FE} to the face images of the speaker. Since the \textit{FE} was trained to minimize the loss between the embedding vector from the \textit{SE} and output vector from the \textit{FE}, SYNTH-AUDIO can be considered as the upper limit of this framework.

In Table \ref{tab:feature_similarity}, we calculate the differences between speech features extracted from generated audio using different methods and the true speech feature ($\rm T_{speech}$). We compare their distance by the $L_{1}$, $L_{2}$ and $Cos$. It can be seen that the speech obtained by FR-PSS is close to the SYNTH-AUDIO. In particular, adding gender prior knowledge can improve performance significantly. For male priors, we only use male speakers for training and testing, and for female priors we only use female speakers for training and testing. Compared with neutral prior experiments, gender prior avoids gender differences, so it gets smaller errors.

Besides, we compare the face features obtained by the \textit{FE} and $\rm T_{speech}$ (see Table \ref{tab:feature_similarity_face}). Results show that by adding prior knowledge, the values of $L_{1}$ and $L_{2}$ decrease dramatically, and the $Cos$ approaches to 1.

%To further evaluate the quality of our proposed FR-PSS, we conduct a gender and age recognition experiment based on the synthesized speech by our FR-PSS and SYNTH-AUDIO. The IFLYTEK REST API \cite{faceplus} is utlized to classify the gender and age of speakers. The dataset we use for test is about 60\% male and 40\% female. Experimental results in Table \ref{tab:gender_accuracy.} show that by adding the speech prior knowledge, the recognition accuracy of both age and gender increases significantly, compared with that using no speech prior.

\subsection{Qualitative Analysis}

\begin{figure}[htb]
	\centering
	\includegraphics[width=0.35\textwidth]{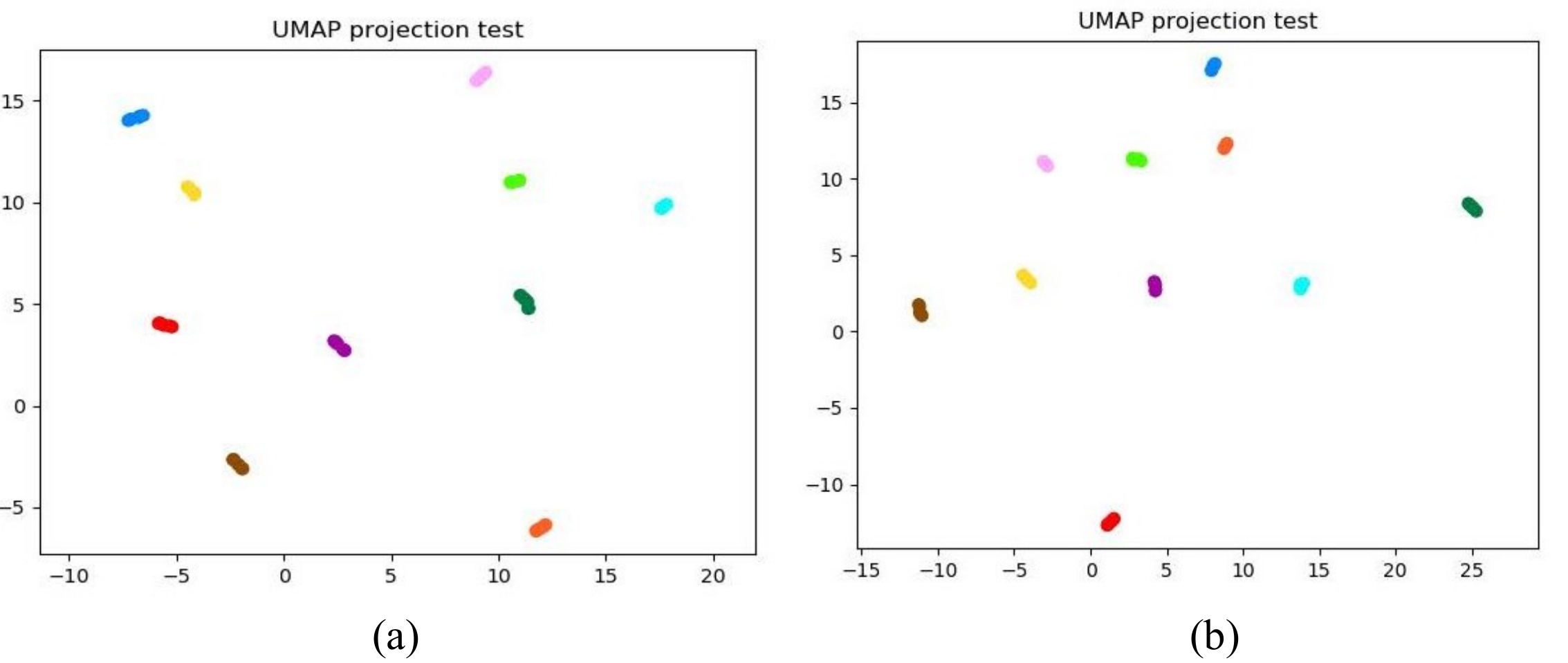}
	\caption{Visualization of speaker embeddings. (a): features of face image. (b): features of synthesized speech by FR-PSS. The same color represents the same person.}
	\label{fig:pipeline3}
\vspace{-0.4cm}
\end{figure}

A visualization of speakers embedding is shown in Fig. \ref{fig:pipeline3} using the U-MAP \cite{2018UMAP}. We randomly select 10 speakers with 5 utterances per speaker. It can be seen that speakers features are well separated in the speaker embedding space for different speakers, either using the face feature or using the synthesized speech feature by our FR-PSS. Besides, different utterances for the same speaker are well clustered. This shows that the speech synthesized by our model can well represent the personalized speech information.
\vspace{-0.6cm}
\begin{table}[htb]
  \caption{Performance of the proposed FR-PSS. $\rm FR-PSS$,  $\rm FR-PSS_{n}$, $\rm FR-PSS_{f}$ and $\rm FR-PSS_{m}$ means FR-PSS with no speech prior, neutral prior, female prior and male prior, respectively. The difference between them and $\rm T_{speech}$ are shown. $\downarrow$ represents the smaller the better, and $\uparrow$ represents the larger the better.}
  \setlength{\tabcolsep}{7mm}
  \label{tab:feature_similarity}
  \centering
  \resizebox{.48\textwidth}{!}{
 \begin{tabular}{ccccccc}
		\hline
		\textbf{Method} &   $L_{1}\downarrow$ & $L_{2}\downarrow$ & $Cos\uparrow$   \\
		\hline
		SYNTH-AUDIO      & 6.251        & 0.664      & 0.887    \\
		$\rm FR-PSS$ & 8.388 & 0.865 & 0.869  \\
		$\rm FR-PSS_{n}$ & {7.769} & {0.805}  & {0.836} \\
		$\rm FR-PSS_{f}$   & 6.963 & 0.785 & 0.881  \\
		$\rm FR-PSS_{m}$  & 7.129 & 0.809 & 0.892  \\
		\hline
	\end{tabular}}
\end{table}
\vspace{-0.5cm}
\begin{table}[htb]
  \caption{Difference between the face features generated by the \textit{FE} and the true speech features.}
  \setlength{\tabcolsep}{7mm}
  \label{tab:feature_similarity_face}
  \centering
\resizebox{.48\textwidth}{!}{
  \begin{tabular}{ccccccc}
		\hline
		\textbf{Method} &   $L_{1}\downarrow$ & $L_{2}\downarrow$ & $Cos\uparrow$   \\
		\hline
		$\rm FR-PSS$      & 9.579       & 0.896      & 0.847    \\
		$\rm FR-PSS_{n}$ & 5.920 & 0.567 & 0.910  \\
		$\rm FR-PSS_{f}$ & {5.462} & {0.509}  & {0.921} \\
		$\rm FR-PSS_{m}$   & 5.613 & 0.516 & 0.943  \\
		\hline
	\end{tabular}}
\end{table}
\vspace{-0.5cm}
\subsection{Subjective Evaluation}
MOS is an assessment experiment based on subjective listening test, and it includes two dimensions: naturalness and similarity. To verify the quality of the speech synthesized by our proposed model, we invite 12 listeners to participate in this evaluation. In the naturalness evaluation, each listener rates the naturalness of 100 speech samples. In the similarity evaluation, each listener is given 20 pairs of samples and is asked to score how similar the synthesized speech is to the real one. All our MOS evaluations are aligned to the absolute category rating scale \cite{1996P}, with rating scores from 1 to 5, and the experimental results are shown in Table \ref{tab:MOS}.
\vspace{-0.5cm}
%That is, to assess how natural the synthesized speech sounds are compared to human real speech sounds, and how similar the synthesized speech sounds are to those of the real speaker. 

\begin{table}[htb!]
    \caption{MOS on subjective listening test.}
    \setlength{\tabcolsep}{7mm}
    \label{tab:MOS}
    \centering
    \resizebox{.48\textwidth}{!}{
    \begin{tabular}{cccc}
    \hline
       \textbf{Method}  & \textbf{Naturalness}$\uparrow$ & \textbf{Similarity}$\uparrow$ \\ 
    \hline
        SYNTH-AUDIO  & 3.83 & 2.43 \\ 
        FR-PSS(proposed)  & 3.60 & 2.12 \\ 
        Face2Speech       & 3.52 & 1.98 \\
    \hline
    \end{tabular}}
\end{table}
\vspace{-0.3cm}
%While the embedding vector of SYNTH-AUDIO is made by applying the \textit{SE} to the audio of the speaker, that of FR-PSS (Face) is generated by applying the \textit{FE} to the face images of the speaker. Since the \textit{FE} was trained to minimize the loss between the embedding vector from the \textit{SE} and output vector from the \textit{FE}, SYNTH-AUDIO can be considered as the upper bound of this framework.

Since our work uses face information to synthesize personalized speech, the MOS score is natural slightly lower than SYNTH-AUDIO which is the upper limit of this framework. However, we extract speech features directly from face images, and a good performance is achieved compared with the model that extract speech features from speech to synthesize personalized audio. And in terms of naturalness and similarity, our proposed method is better than Face2Speech. Samples are available at https://github.com/hxs123hxs/FR-PSS\_example.
% Our proposed model is about 0.2 lower in naturalness than SYNTH-AUDIO, because speaker identity is not as well separated from face image as it is from speech. In terms of similarity, it can be seen from Table \ref{tab:gender_accuracy.} that FR-PSS can basically show the correct gender and age range of the speaker.The score of our proposed model dropped slightly, suggesting that probably some subtle differences, such as those related to characteristic prosody, are not represented accurately enough. 
\vspace{-0.5cm}
\section{Conclusion}
\vspace{-0.3cm}
In this work, we innovatively propose the FR-PSS model to extract personalized speech features from human faces and synthesize personalized speech using neural vocoder. By designing three speech priors, a residual-guided strategy is introduced to guide the face feature to approach the true speech feature as much as possible. Experimental results show that our FR-PSS achieves a satisfying performance compared with the method that synthesizes personalized speech using a speech-derived embedding vector. In the future, we will investigate a multi-task framework to further improve the efficiency of the personalized speech synthesis.
\vspace{-0.3cm}
\section{Acknowledgements}
\vspace{-0.15cm}
This work was supported by the National Natural Science Foundation of China(No.61977049), National Natural Science Foundation of China(No.62101351) and the Tianjin Key Laboratory of Advanced Networking.

%\vfill\pagebreak

% References should be produced using the bibtex program from suitable
% BiBTeX files (here: strings, refs, manuals). The IEEEbib.bst bibliography
% style file from IEEE produces unsorted bibliography list.
% -------------------------------------------------------------------------
\bibliographystyle{IEEEbib}
\bibliography{refs}

\end{document}